\documentclass[epj]{svjour}
\usepackage{epsfig}

\usepackage{graphics}
\def\f2{$F_2^{\gamma}(x,Q^2)$}

\begin{document}

\title{Measurements of the Photon Structure Function at LEP}
\author{A. De Roeck
}                     
%
%
\institute{CERN, 1211 Geneva 23, Switzerland}
\date{}
%
\abstract{
In this contribution we discuss recent measurements by the LEP experiments
on photon structure functions.
}
%
\maketitle

\vspace{0.8cm}
\section{Introduction}
\label{intro}
Photon structure functions are traditionally  measured in $e\gamma$
scattering at electron-positron colliders.
One of the electrons emits a 
photon which is almost on mass shell, and which is probed by a photon
with virtuality $Q^2$ emitted by the second electron. The latter electron
will be scattered within the acceptance of the detectors if the
polar angle is larger than approximately 25 mrad,
while the first electron will go undetected in the beam-pipe.
Hence the signature is a 
so called single-tag event. For a recent review of the 
photon structure see~\cite{nisius}.

Recent new data  include measurements of 
\f2, where $x$ is the Bjorken-$x$ value of the parton in the photon,
 from ALEPH and DELPHI; the charm structure
function of the photon
from OPAL;  a first measurement of the electron structure function 
by DELPHI. New parton density parametrizations have been extracted 
using the (almost) complete data sets on \f2, and an interesting extraction of 
$\alpha_s$ has been reported.

\section{New ALEPH data on {\boldmath \f2}}
\begin{figure}[htb]
\centering
\epsfig{file=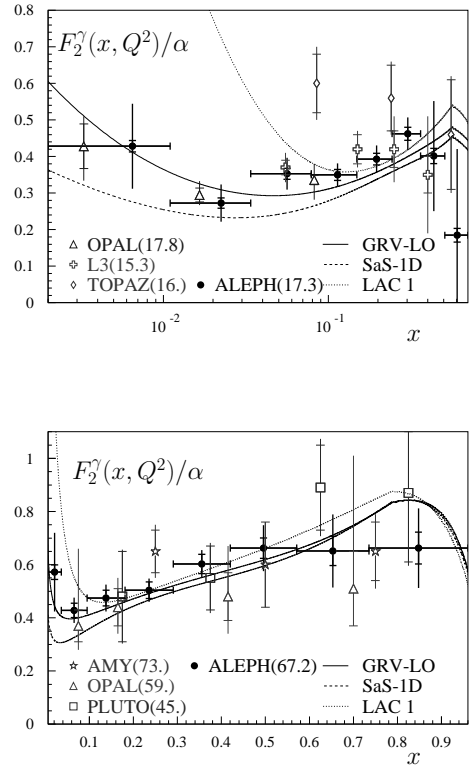,bbllx=60pt,bblly=170pt,bburx=430pt,bbury=730pt,height=11.0cm}
\caption{Values of \f2/$\alpha$ from ALEPH, compared to previous measurements
and predictions of photon PDFs.}
\label{aleph}
\end{figure}

ALEPH~\cite{aleph} 
reports a measurement of \f2 based on 584.4 pb$^{-1}$, in the medium 
$Q^2$ range: at 17.3 GeV$^2$ and 67.3 GeV$^2$. The Tikhonov unfolding 
procedure was used to extract the data,
and the results are shown in Fig.~\ref{aleph}. The data are compatible 
with earlier measurements, but are more precise.
In contrast to  the proton, the structure 
function of the photon is predicted to rise linearly with the 
logarithm of the momentum transfer $Q^2$, and to increase with
increasing Bjorken-$x$~\cite{gg_zerwas}.
The ALEPH data  allows to check the rise of \f2 with increasing
$Q^2$:
$F_2^{\gamma}(0.1< x <0.5, <Q^2>=17.3\, $GeV$^2) =0.41\pm0.01(stat.) \pm 0.08
(sys.)$ and 
$F_2^{\gamma}(0.1< x <0.7, <Q^2>=67.2\, $GeV$^2) =0.52\pm0.01(stat.) \pm 0.06
(sys.)$.

\section{New DELPHI data on {\boldmath \f2}}
\begin{figure}[htb]
\centering
 \epsfig{file=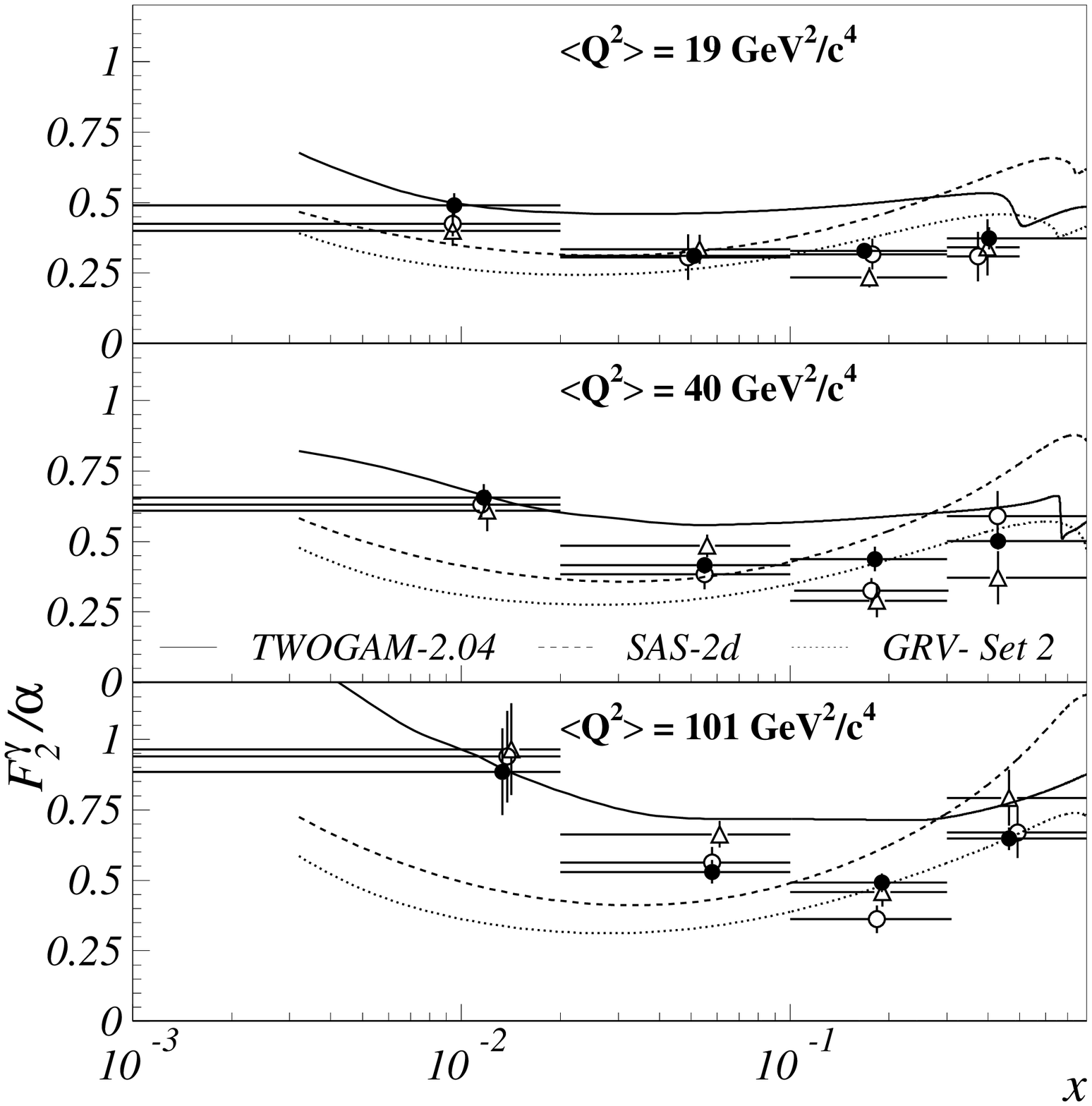,bbllx=0pt,bblly=0pt,bburx=590pt,bbury=590pt,height=5.0cm,clip=}
 \epsfig{file=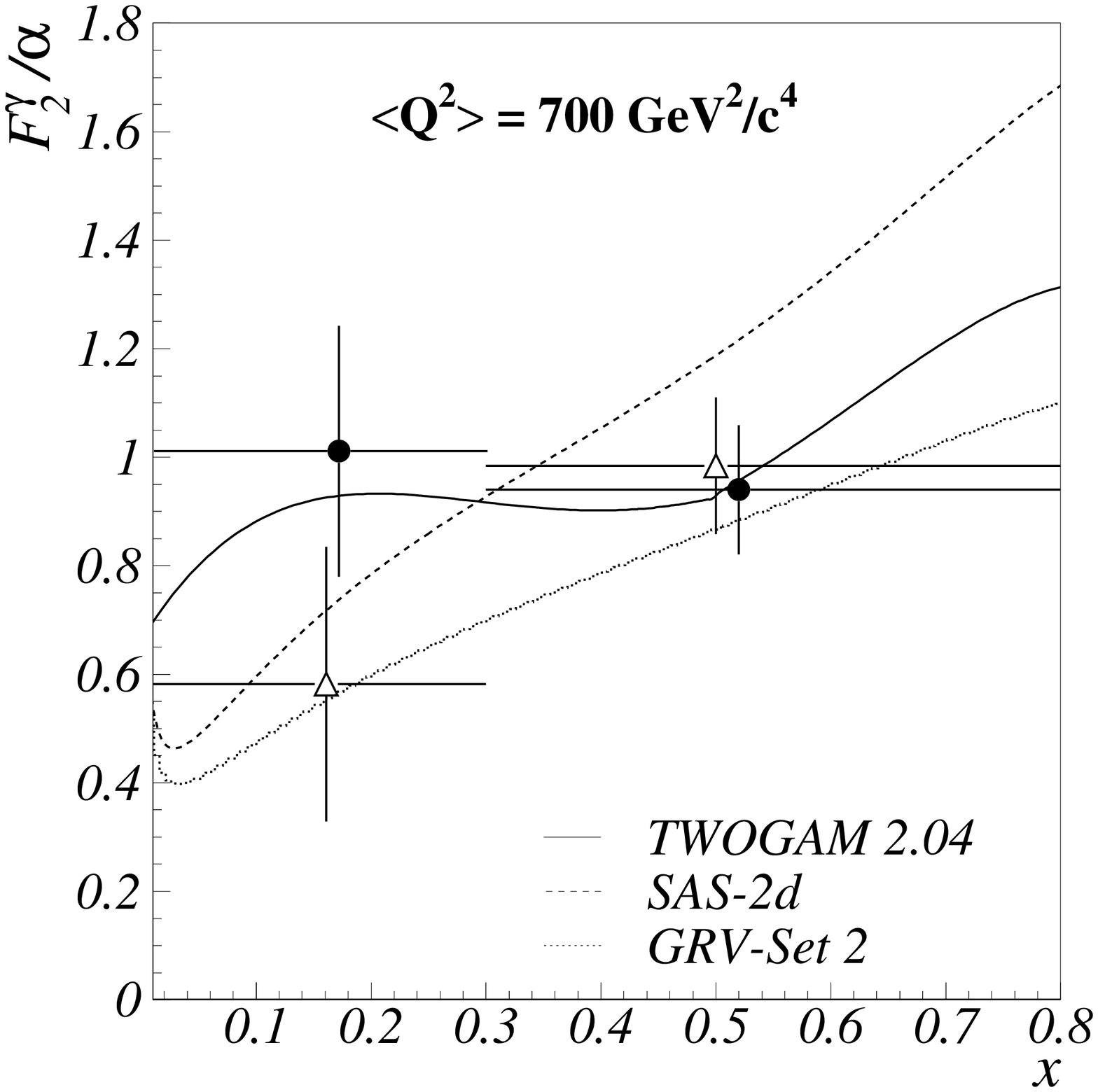,bbllx=0pt,bblly=0pt,bburx=590pt,bbury=590pt,height=5.0cm,clip=}
\caption{\f2/$\alpha$ measurements
 at different $Q^2$ values, using LEP2 data.
The results are extracted from data using TWOGAM (black points), PYTHIA (open 
triangles) and PHOJET(open circles) and are compared to model predictions.}
\label{delphi_2}
\end{figure}

DELPHI~\cite{delphi-1} presents new LEP1 and LEP2 data analyses based on 
78 pb$^{-1}$ and 548 pb$^{-1}$ respectively. No unfolding procedure is used but
the different cross section components are fitted to the data based on 
hadronic models.
The LEP2 data  are shown in 
 Fig.~\ref{delphi_2}. DELPHI
chooses to present different \f2 values calculated/corrected with different
hadronic models, but do not give a single measurement with a total error
including the  hadronic uncertainty.
 Hence it is somewhat difficult to compare these measurements 
with results from other experiments.

\section{OPAL data on {\boldmath $F_{2,c}^{\gamma}(x,Q^2)$}} 
OPAL reported end of last year on a new measurement of the 
charm content of \f2, namely $F_{2,c}^{\gamma}(x,Q^2)$\cite{opal}, 
which depends directly on the
gluon content of the photon.
The complete luminosity collected by OPAL  during the years 
1997-2000 was used, namely 654.1 pb$^{-1}$. The measurement is made 
in the $Q^2$ region of $5 < Q^2< 100 $ GeV$^2$.
The decays $D^{*+}\rightarrow D^0\pi^+
\rightarrow K^-\pi^+\pi^+ $ and charge conjugates, have been used.
After selection cuts, exploiting the small mass difference between  the 
$D^*$ and the $D$
decay,  $55.3\pm11.0$ signal events are selected.
Divided in two bins in $x$ gives 
$23.6\pm 7.4 $ events for
$x < 0.1$
and
$31.4\pm 8.1 $ events for 
$x > 0.1$.

Converting the number of events to a structure function measurement
 and
comparing this
with QCD calculations, one finds that the high-$x$ region is well
described but the low-$x$ data is above the prediction.
 In the high-$x$ region the charm structure function is dominated by
the pointlike part of the cross section.
If one subtracts the NLO pointlike part from the low-$x$ data then the 
resulting measurement of the hadronic part of the cross section 
for $0.0014<x<0.1$ gives
$34.5\pm14.3\pm 6.9$ pb
with a theory prediction of $7.7^{+2.2}_{-1.6}$ pb, i.e. the 
low-$x$ excess
is about $2\sigma$.

A further improvement of this measurement can only be achieved if also the 
other LEP experiments will have a go at it.

\begin{figure}[htb]
\centering
 \epsfig{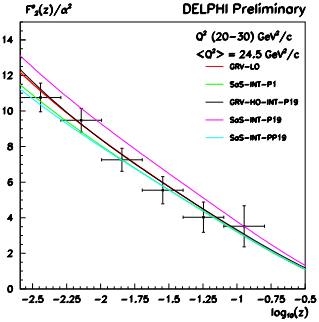}
\caption{The electron structure function averaged in the region of $Q^2$
= 20-30 GeV$^2$.}
\label{delphi_3}
\end{figure}

\section{Electron structure function}
A few years ago it was argued~\cite{szwed} that one could circumvent some
of the dominant systematics of present photon structure function measurements,
namely the necessity to measure and use the hadronic final state to reconstruct
the kinematic variable $x$,
by making a measurement of the electron structure instead.
Here the 
emitted real photon is considered as part of the structure of the electron. 
The kinematics of the process can be reconstructed from 
 the scattered electron, and no unfolding procedure is necessary. 
Furthermore no correction due to the 
small but finite virtuality of the target photon is needed. On the downside
the rapid falling photon flux with increasing photon energy
is now absorbed in the measurement (instead
of being factorized out as for the photon structure function case) 
and obscures the 
sensitivity to the QCD dynamics of the photon structure. Furthermore the
radiative corrections can become quite large.

DELPHI has presented a first preliminary measurement~\cite{estruc}, shown
in Fig.~\ref{delphi_3}.
The electron structure function falls rapidly with increasing $z$ (= the 
Bjorken-$x$ w.r.t. the electron)  as expected.
The measurements are consistent with the \f2 results, but within the
large error bars no significant increased sensitivity to the underlying
dynamics is seen: the predictions of the different models are much closer
to each other than in the case of the photon structure function.
It constitutes however an important cross check of the photon structure
function measurement, and it is noted~\cite{estruc} that the statistical
uncertainties on this measurement may be better understood.

\begin{figure}[htb]
\centering
 \epsfig{file=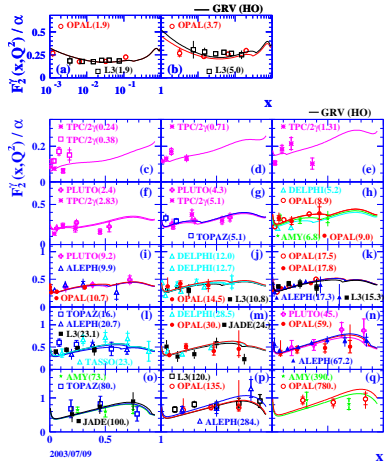,bbllx=60pt,bblly=200pt,bburx=520pt,bbury=800pt,height=6.5cm}
\caption{Summary of measurements of the hadronic structure function
\f2, compared with the GRV-HO prediction~\cite{nisius}.}
\label{total1}
\end{figure} 

\begin{figure}[htb]
\centering
 \epsfig{file=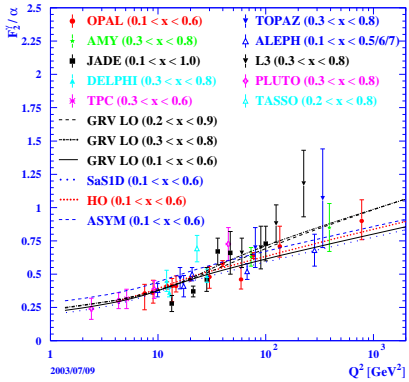,bbllx=0pt,bblly=0pt,bburx=550pt,bbury=530pt,height=5.5cm}
\caption{Measurements of the $Q^2$ evolution of \f2 compared to a linear fit
of the type $a+b\ln(Q^2/\Lambda^2)$ with $\Lambda = 0.2$ GeV (dashed line)
and the $1 \sigma$ errors of the fit (dotted line). In addition photon PDFs
are  shown~\cite{nisius}.}
\label{total2}
\end{figure}

\section{Summary of the world data}
A summary of the world data is shown in Figs~\ref{total1}
and ~\ref{total2}. The first Figure shows the data in different $Q^2$ bins 
as function of $x$. It contains in total over 50 measurements in the
 kinematical range $0.001< x < 0.9$ and 
$1.9 < Q^2 < 780$ GeV$^2$.

In Fig.~\ref{total2} the data are shown in different $x$ bins as function of 
$Q^2$. A fit of the form \f2 =$a+b\ln(Q^2/\Lambda^2)$ gives $b$ values
of $0.061\pm0.003, 0.095\pm0.008$ and $0.135\pm 0.013$, for 
$x$ ranges of $0.01-0.1, 0.2-0.3 $ and $ 0.4-0.6$ respectively, and
$\Lambda = 0.2 $ GeV. Hence there is a significant increase of the slope with 
increasing $x$.

\section{Parton distributions and {\boldmath $\alpha_s$}}
Using all \f2 data measured at LEP, apart from the new data reported
in this paper, recently new parton distributions and an extraction of the 
strong coupling constant $\alpha_s$ have been reported.

LO parton distributions were reported in~\cite{cornet}. These PDFs were
radiatively generated, and use the ACOT (and FFNS) heavy flavour scheme.
NLO densities are in progress.

Fits to the \f2 data to extract $\alpha_s$ were reported in \cite{albino}.
Two different extractions have been made.
The first uses data with $x> 0.45, Q^2> 59 $ GeV$^2$, and $\alpha_s$ is
fitted directly from its logarithmic asymptotic behaviour.
The result (NLO/$\overline{MS}$) is 
$\alpha_s = 0.1183\pm 0.0050(exp)^{+0.0029}_{-0.0028}(theo)$.
The second fit uses all data but makes a 5 parameter fit,
one of which is $\alpha_s$. The result of this fit is
 $\alpha_s = 0.1198\pm0.0028(exp)^{+0.0034}_{-0.0046}(theo)$.
The results are consistent with each other and the precision is interesting
in view of the total precision of the world data 
$\alpha_s = 0.1172\pm0.0020$~\cite{pdg}.

\section{Outlook and conclusion}
In the near future what can we still expect from LEP?
OPAL works on a low-$x$ analysis for \f2 and $F_2^{e}(z,Q^2)$ using the 
complete statistics; L3 plans an analysis in the full kinematical plane
using the complete LEP statistics, and ALEPH may possibly perform 
an analysis with the full data sample at high $Q^2$.

On the longer term, an extension of the present kinematic region and 
 improved quality of the data can be expected only at 
a future linear collider~\cite{nisius} or photon collider~\cite{deroeck}.
At a photon collider the coverage and precision can be of 
similar quality as the one of today's proton measurements.

In summary the structure of the photon is now measured in the 
kinematical range of $0.001< x < 0.9$ and 
$1.9 < Q^2 < 780$ GeV$^2$. The precision has been improving over
the years. The charm structure function measurement can become significant
if the data of all experiments will be analysed and combined. 
A first extraction of the 
electron structure function has been presented.

{\bf Acknowledgement}

Thanks to A. Finch, C. Mariotti, B. Muryn, R. Nisius, T. Szumlack, T. Wengler

\end{document}